# Mesoscopic chaos mediated by Drude electron-hole plasma in silicon optomechanical oscillators

Jiagui Wu[1,2], Shu-Wei Huang[1], Yongjun Huang[1], Hao Zhou[1], Jinghui Yang[1], Jia-Ming Liu[3], Mingbin Yu[4], Guoqiang Lo[4], Dim-Lee Kwong[4], Shukai Duan[2,*], and Chee Wei Wong[1,*]

[1] Fang Lu Mesoscopic Optics and Quantum Electronics Laboratory, University of California Los Angeles, CA 90095

[2] College of Electronic and Information Engineering, Southwest University, Chongqing, China 400715

[3] Electrical Engineering, University of California Los Angeles, CA 90095

[4] Institute of Microelectronics, A*STAR, Singapore 117865

* Email: duansk@swu.edu.cn; cheewei.wong@ucla.edu

**Chaos has revolutionized the field of nonlinear science and stimulated foundational studies from neural networks, extreme event statistics, to physics of electron transport. Recent studies in cavity optomechanics provide a new platform to uncover quintessential architectures of chaos generation and the underlying physics. Here we report the generation of dynamical chaos in silicon-based monolithic optomechanical oscillators, enabled by the strong and coupled nonlinearities of two-photon-absorption induced Drude electron-hole plasma. Deterministic chaotic oscillation is achieved, and statistical and entropic characterization quantifies the chaos complexity at 60 fJ intracavity energies. The correlation dimension $D_2$ is determined at 1.67 for the chaotic attractor, along with maximal Lyapunov exponent rate about 2.94× the fundamental optomechanical oscillation for fast adjacent trajectory divergence. Nonlinear dynamical maps demonstrate the subharmonics, bifurcations, and stable regimes, along with distinct transitional routes into chaos. This provides a CMOS-compatible and scalable architecture for understanding complex dynamics on the mesoscopic scale.**

Investigation of chaos and the associated nonlinear dynamics has spurred fundamental progress of science and technology. It brought new perspectives in a multitude of fields spanning from recurrent neural networks [1], relativistic billiards-like electron transport [2], fractal space and time [3] to self-organization in natural sciences [4], amongst others. Chaos in optical systems has emerged and drawn much attention due to its unique features and broad applications, including chaos-based synchronized secure optical communications [5-7], high-performance light detection and ranging [8] and ultrafast physical random bit generation [9]. Studies of chaos generation in III-V laser components have further shown progress in harnessing the broadband carriers in both the near infrared and the mid-infrared wavelength ranges [10-17], although the challenges of monolithic integration and circumventing the seemingly universal requirement of external perturbations remains to be solved.

Concurrently, significant efforts in nanofabrication technology and cavity optomechanics have led to the demonstration of regenerative oscillation in mesoscopic resonators [18-21]. Driven by centrifugal radiation pressure, optomechanical chaotic quivering was experimentally observed in toroidal whispering-gallery-mode (WGM) microcavities [22]. Recently, in a toroidal WGM microcavity, stochastic resonance and chaos are transferred between two optical fields [23], with the chaotic physical basis through a strong nonlinear optical Kerr response from the nonlinear optical-mechanical modal coupling. This is complemented by recent theoretical studies on chaos including electro-optomechanical systems and potential routes into chaos [24, 25]. Here, we couple the prior single optomechanical basis with a second basis – that of electron-hole plasma oscillations in the same cavity – to deterministically generate dynamical chaos in a silicon photonic crystal (PhC) cavity. Differing from the prior studies, the silicon experimental platform enables electron-hole plasma dynamical generation, destabilizing the system dynamics, and provides a route for chip-scale planar electronic-photonic integration. Our photonic crystal implementation is based on a slot-type optomechanical (OM) cavity with sub-wavelength [$\approx 0.051(\lambda/n_{air})^3$] modal volumes $V$, and high quality factor ($Q$) to $V$ ratios [26, 27]. This provides strong optical gradient oscillation [26, 28] to achieve operating intracavity energies at approximately 60 fJ, and for near-single-mode operation. Our two-oscillator OM cavity is designed with comparable dynamical oscillation timescales between Drude electron-hole plasma and radiation pressure optomechanics -- this allows the chaotic attractors and unique trajectories to be uncovered for the first time. We present the statistical and entropic characteristics of the nonlinear dynamical regimes, and illustrate the transition routes into and out of chaos. Our first-principles numerical modeling, including coupled oscillations in seemingly unrelated degrees-of-freedom - two-photon-induced free-carrier and thermal dynamics with radiation pressure dynamics - capture the experimental observations, the multi-period orbits, and the trajectory divergence into chaotic states.

**Results**

**Experimental observation of chaos.** Figures 1a shows the scanning electron micrograph (SEM) of the slot-type optomechanical photonic crystal cavity mediated by Drude electron-hole plasma examined in this study. The air-bridged photonic crystal cavity is introduced with shifted-centre air holes that are shifted by 15 nm, 10 nm, and 5 nm, respectively, as shown in Figure 1b. The width-modulated line-defect photonic crystal cavity design has a total quality factor $Q$ of 54,300 (Figure 1c) and a sub-wavelength modal volume of $0.051(\lambda/n_{air})^3$ (Figure 1b inset) at the 1572.8 nm resonance wavelength ($\lambda_o$, with the effective mode index $n$). The optomechanical cavity consists of two (16.0 μm × 5.5 μm × 250 nm) micromechanical photonic crystal slabs, separated by a 120 nm slot width across the photonic crystal line defect. The in-plane mechanical mode has a 112 MHz fundamental resonance and, when driven into the regenerative oscillation regime, has a narrow sub-15-Hz linewidth at ambient pressure and room temperature [29]. The large optical field gradient from the tight slot cavity photon confinement enables a large coherent optomechanical coupling strength $g_0$ of ~ 690 kHz (detailed in Supplementary Note 4 ), resulting in low-threshold optomechanical oscillation (OMO) [26-29]. Concurrently, on the same cavity, strong nonlinearities such as two-photon absorption, free-carrier and thermo-optic dynamical effects lead to modulation of the intracavity field [30]. Note the characteristic timescales of the OMO and the photonic crystal carrier dynamics are made comparable, through our designed mechanical modes and intrinsic free-carrier diffusion times, enabling the coupled equations of motion to have sufficient overlap and degrees of freedom for chaos generation.

    Figure 1d depicts the transition into chaos as the pump detuning to the cavity resonance $\Delta$ (= $\lambda_L$ - $\lambda_o$, where $\lambda_L$ is the injection light wavelength) is scanned from 0.2 nm to 4.2 nm with the injection power fixed at 1.26 mW (detailed in Methods). The chaos region as well as the

associated dynamical transitional states can be identified. First, a stable pure fundamental OMO at 112 MHz is observed at the beginning of the detuning drive. With increased detuning, aperiodic and sub-oscillatory structures emerges when $\Delta$ is set in the range of 1.2 to 2.0 nm. Unstable pulses (USP) occur first, before the system is driven into a series of stable sub-harmonic pulse states such as the $f_{omo}/4$ states (oscillation period being four times the OMO period), the $f_{omo}/3$ states, and the $f_{omo}/2$ states respectively. For detuning $\Delta$ between 2.0 and 2.33 nm, the system observes a chaos region characterized by both a broadband radio frequency (RF) spectrum and an intricate phase portrait. For detuning $\Delta$ greater than 2.33 nm, the system is driven to exit the chaos region by evolving into a $f_{omo}/2$ state ($\Delta$ = 2.33 to 3.2 nm) before cumulating into a self-induced optical modulation (SOM) state ($\Delta$ = 3.2 to 4.2 nm) [30, 31]. Of note, the oscillation period of SOM (about 13 ns to 17 ns), mainly determined by the Drude plasma effect and thermal dissipation rate, is comparable with that of OMO (about 9 ns). The close oscillation frequencies of SOM and OMO facilitate their effective interaction in PhC nanocavity and the occurrence of chaos [4, 18].

Figure 2 shows an example chaotic oscillation in the temporal domain and its RF frequency spectrum with the recorded raw temporal waveform shown in Figure 2a, illustrating the irregular and intricate fluctuations. Figure 2b presents the phase portrait of chaos in a two-dimensional plane spanned by the power of temporal waveform (*P*, horizontal axis) and its first derivative ($\sigma$, vertical axis) [32]. The reconstructed trajectory is useful for illustrating the complex geometrical and topological structure of the strange attractor, showing the local instability, yet global stable nature, of a chaos structure [32]. In order to reveal the topological structure of chaos attractors, a state-space procedure is implemented to average the temporal waveform points in a *m*-dimensional embedded space [32] (detailed in Supplementary Note 1) by removing stochastic noise from the recorded raw data. The noise removal enables a clear depiction of the topological structure of attractor and is also useful for the estimation of correlation dimension and Kolmogorov entropy, the most commonly used measures of the strangeness of chaotic attractor and the randomness of chaos [33-36]. Furthermore, Figure 2c shows the corresponding RF spectrum, where the signal distributes broadly and extends up to the cutoff frequency of measurement instrumentation, showing a hallmark spectral feature of chaos.

Figure 3 illustrates the detailed properties of several different dynamical states, including RF spectra, temporal waveforms, and phase portraits. First, Figure 3a shows the frequency and temporal characteristics of the $f_{omo}/2$ state. We observe three characteristic features of the $f_{omo}/2$ state: distinct $f_{omo}/2$ components in the RF spectrum (Figure 3a), pulses with period ($\approx$ 17.8 ns) at two times the OMO period ($\approx$ 8.9 ns) in the temporal waveform (Figure 3b), and clear limit cycle [37] features in the phase portrait (Figure 3c). Similarly, Figures 3d-3f and 3g-3i show the frequency spectra, the temporal waveforms at a third and a quarter of the fundamental oscillation, and the corresponding limit cycle phase portraits of the transitional $f_{omo}/3$ and $f_{omo}/4$ states, respectively. We note the satellite bumps next to the main peaks in the temporal waveforms; they represent the relatively weak OMO fundamental oscillations. Figures 3j and 3k next show the frequency and temporal features of the chaos state, where a broadband spectrum and a fluctuating temporal waveform are observed. In the phase portrait (Figure 3l), the trajectory evolves intricately and scatters widely in phase space, being quite different from other periodical dynamics. With this slot cavity and at 1.26 mW injection power (about 60 fJ intracavity energy), the specific transition route is OMO - USP - $f_{omo}/4$ - $f_{omo}/3$ - $f_{omo}/2$ - chaos - $f_{omo}/2$ - SOM, exhibiting a clear sub-harmonic route to chaos. The complete set of routing states into/out of chaos is detailed in Supplementary Note 2.

**Dynamical characterization of chaos**. Statistical analysis is next performed to uncover the detailed dynamical properties of the chaotic states. A three-dimensional phase space is constructed in Figure 4a, in a volumetric space spanned by the power (*P*), the first time derivative of *P* ($\sigma$) and the second time derivative of *P* ($\xi$). The green curves are the projections of the trajectory onto each of the three phase planes, showing the geometric structures. Three statistical measures, Lyapunov exponents (LEs), correlation dimension, and Kolmogorov entropy, are commonly employed to illustrate and characterize the dynamical properties of chaos [32-38]. Details of these measures are provided in Supplementary Note 1. LEs, which describe the divergence rate of nearby attractor trajectories, are the most widely employed criteria in defining chaos [33]. In Figure 4b, we show the calculated LEs, converging to values $\lambda_1 \approx$ 0.329 ns$^{-1}$, $\lambda_2 \approx$ -0.087 ns$^{-1}$ and $\lambda_3 \approx$ -0.946 ns$^{-1}$ respectively, or equivalently when expressed on the intrinsic optomechanical photonic crystal cavity timescale ($\tau_{omo} = f_{omo}^{-1} \approx 8.9$ ns) $\lambda_1 \approx 2.94\tau_{omo}^{-1}$, $\lambda_2 \approx -0.78\tau_{omo}^{-1}$ and $\lambda_3 \approx -8.45\tau_{omo}^{-1}$. The maximal LE is positive, illustrating a fast divergence rate between adjacent orbits and indicating that the system is chaotic [32, 33]. We further analyze the correlation dimension $D_2$:

$$D_2 = \lim_{\substack{D \to \infty \\ r \to 0}} \frac{d \ln(C_D(r))}{d \ln(r)} \qquad (1)$$

where $C_D$ is the correlation integral of vector size $D$ in an $r$ radius sphere and $d$ is the Euclidian norm distance [36]. A conservative estimate of the attractor correlation dimension is implemented through the Grassberger-Procaccia (G-P) algorithm [36, 38] as detailed in Supplementary Note 1. As shown in Figure 4c, the correlation integrals $C_D$ vary with sphere radius $r$. In Figure 4d, the plot of correlation integral slope versus sphere radius $r$ is obtained by extracting the slope from Figure 4c. A clear plateau of the correlation integral slope is observed, supporting the estimated value of $D_2$ at about 1.67 ($D_2 \approx 2.0$ without noise filtering). The correlation dimension $D_2$ highlights the fractal dimensionality of the attractor and demonstrates the strangeness of the complex geometrical structure [34]. We note that this $D_2$ value is already higher than that of several canonical chaos structures such as the Hénon map (at 1.21), the logistic map (at 0.5), and the Kaplan-Yorke map (at 1.42), and is even close to that of Lorenz chaos (at 2.05) [36].

Furthermore the waveform unpredictability can be characterized by the second-order Renyi approximation of the Kolmogorov entropy $K_2$:

$$K_2 = \lim_{\substack{D \to \infty \\ r \to 0}} \frac{1}{\tau} \ln\left(\frac{C_D(r)}{C_{D+1}(r)}\right) \qquad (2)$$

where $\tau$ is the time series sampling rate], a measurement of the system uncertainty and a sufficient condition for chaos [38]. A positive $K_2$ is characteristic of a chaotic system, while a completely ordered system and a totally random system will have $K_2 = 0$ and $K_2 = \infty$ respectively. With the G-P algorithm, $K_2$ is calculated as $\approx 0.17$ ns$^{-1}$ or expressed equivalently as $\approx 1.52\tau_{omo}^{-1}$, representing that the mean divergence rate of the orbit section (with adjoining point pairs in the phase space) is rapid within 1.52 times the fundamental OMO period. It characterizes the gross expansion of the original adjacent states on the attractor [38] and, therefore, indicates the significant unpredictability in the dynamical process of such solid-state systems.

**Theoretical simulation of chaos**. To further support the physical observations, we model the dynamics of the optomechanical photonic crystal cavity system under the time-domain nonlinear coupled mode formalism, taking into account the OMO oscillation [21], two-photon absorption [31], free-carrier and thermo-optic dynamics [30, 31]:

$$\frac{d^2x}{dt^2} + \Gamma_m \frac{dx}{dt} + \Omega_m^2 x(t) = \frac{g_0}{\omega_0}\sqrt{\frac{2\Omega_m}{\hbar m_{eff}}}|A(t)|^2 \tag{3}$$

$$\frac{dA}{dt} = i\left(-g_0\sqrt{\frac{2m_{eff}\Omega_m}{\hbar}}x(t) + \frac{\omega_0}{n_{Si}}\left(\frac{dn_{Si}}{dT}\Delta T(t) + \frac{dn_{Si}}{dN}N(t)\right) + \delta\omega\right)A(t)$$

$$-\frac{1}{2}(\gamma_i + \gamma_e + \frac{\Gamma_{TPA}\beta_{Si}c^2}{V_{TPA}n_g^2}|A(t)|^2 + \frac{\sigma_{Si}cN(t)}{n_g})A(t) + \sqrt{\gamma_e P_{in}} \tag{4}$$

$$\frac{dN}{dt} = -\frac{N(t)}{\tau_{fc}} + \frac{\Gamma_{FCA}\beta_{Si}c^2}{2\hbar\omega_0 n_g^2 V_{FCA}^2}|A(t)|^4 \tag{5}$$

$$\frac{d\Delta T}{dt} = -\frac{\Delta T(t)}{\tau_{th}} + \frac{\Gamma_{PhC}}{\rho_{Si}c_p V_{PhC}}\left(\gamma_i + \frac{\Gamma_{TPA}\beta_{Si}c^2}{V_{TPA}n_g^2}|A(t)|^2 + \frac{\sigma_{Si}cN(t)}{n_g}\right)|A(t)|^2 \tag{6}$$

where $x$, $A$, $N$ and $\Delta T$ represent respectively the motional displacement, the intracavity **E**-field amplitude, the free-carrier density, and the cavity temperature variation. $\delta\omega = \omega_L - \omega_0$ is the detuning between injection light $\omega_L$ and photonic crystal cavity resonance $\omega_0$, and $P_{in}$ is the injected optical power (detailed in Supplementary Note 3, Supplementary Table 1). Equation 3 describes the optically-driven damped mechanical harmonic oscillation, with self-sustained OMO oscillations when pumped above threshold. The mechanical oscillations then in turn result in modulation of the intracavity optical field (first term on the right-hand side of Equation 4). On the other hand, the plasma induced thermal-optic effect and free-carrier dispersion (FCD) in the cavity (second and third terms on the right-hand side of Equation 4) lead to another amplitude modulation of the intracavity field. Here, the high-density Drude plasma is generated by the strong two-photon absorption (TPA) in silicon (Equation 5). , With the increased intracavity power, the FCD effect leads to blue-shifts of the cavity resonance while the free-carrier absorption induced thermo-optic effect results in red-shifts of the cavity resonance. The dynamical interplay between these two effects results in the regenerative SOM [30, 31]. The mechanism is detailed in Supplementary Figure 6 and Supplementary Note 6. We note that our PhC design ensures the characteristic timescales of the SOM and OMO oscillations are on the same order of magnitude (Supplementary Figure 8), strengthening the effective inter-oscillator coupling. The coexistence of OMO and SOM mechanisms adds extra degrees of freedom to the dynamic space of system and results in increased susceptibility to destabilization (detailed in Supplementary Note 2) [16, 18, 21].When the drive power is between the SOM and OMO thresholds, TPA-associated amplitude modulations disrupt the OMO rhythm, breaking the closed OMO limit cycles and creating the non-repeating chaotic oscillations. On the other hand, if the frequency ratio between OMO and SOM is close to a rational value, they will lock each other based on the harmonic frequency locking phenomena [39, 40]. Consequently, different sub-harmonic $f_{omo}$ states are also observed in Figure 3. Effects of the Drude free-carrier plasma, detuning $\delta\omega$, optomechanical coupling strength $g_0$, and injected drive power $P_{in}$ on the chaotic transitions and routes are detailed in Supplementary Note 4 to 7.

Figure 4e shows the dynamical distribution map simulated numerically and parametrically with the normalized detuning $\delta\omega/\gamma_i$ versus injection power $P_{in}$, where $\gamma_i$ is the intrinsic cavity linewidth from linear losses. The various regimes are denoted with different colours, and rigorously identified through entropic analysis of the temporal waveform uncertainty and periodicity of the Fourier spectrum. The temporal waveforms are often strongly periodic in the limit cycle states (such as OMO and USP) and have low entropy (indicated by the darker colours), while the chaotic oscillation has a significant uncertainty and high entropy (indicated by the brighter colours). In Figure 4e, the crescent-shaped region (in bright orange) indicates the parametric conditions of the complex chaos state. Around this region, there are rich transitional dynamics related to chaos, thereby enabling different routes into or out of chaos with different parameter scanning approaches. When the pump power is 1.26 mW, the numerical model predicts a bifurcation transition to chaos via states OMO - USP - $f_{omo}$/3 - $f_{omo}$/2 - chaos - SOM as a function of detuning, in a qualitative agreement with the experimental observations. Of note, the system of coupled equations does not involve any initial noise terms, illustrative of the deterministic nature of the obtained chaotic solutions.

**Discussion**

We demonstrate chaos generation in mesoscopic silicon optomechanics, achieved through single-cavity coupled oscillations between radiation-pressure- and two-photon-induced free-carrier dynamics. Chaos generation is observed at 60 fJ intracavity energies, with a correlation dimension $D_2$ determined at approximately 1.67. The maximal Lyapunov exponent rate is measured at 2.94 times the fundamental OMO frequency, and the second-order Renyi estimate of the Kolmogorov entropy $K_2$ is determined at 1.9 times the fundamental OMO period – both showing fast adjacent trajectory divergence into the chaotic states. Furthermore, we route the chaos through unstable states and fractional subharmonics, tuned deterministically through the drive-laser detuning and intracavity energies. These observations set the path towards synchronized mesoscopic chaos generators for science of nonlinear dynamics and potential applications in secure and sensing application, in light of recent works about gigahertz optomechanical oscillation [41] and synchronization of coupled optomechanical oscillators [42].

**Methods**

**Device design and fabrication:** The optomechanical photonic crystal cavity is fabricated with a CMOS-compatible process on 8-inch silicon wafers at the foundry, using 248 nm deep-ultraviolet lithography and reactive ion etching on 250 nm thickness silicon-on-insulator wafers. To realize the critical 120 nm slot width, the resist profile is patterned with a 185 nm slot line width, then transferred into a sloped oxide etch. The resulting bottom 120 nm oxide gap is etched into the silicon device layer through tight process control. Multiple planarization steps enable high-yield of the multi-step optomechanical photonic crystal fabrication. The optical input/output couplers are realized with silicon inverse tapers and oxide overcladding coupler waveguides. The optomechanical photonic crystal cavities are released by timed buffered oxide etch of the undercladding oxide.

**Measurement setup:** The drive laser is a tunable Santec TSL-510C laser (1510 nm to 1630 nm), which is also used to measure the optical transmission spectra. The drive laser is first amplified by a C-band erbium-doped fibre amplifier and then injected into the slot-type photonic crystal cavity with a coupling lens placed on an adjustable 25-nm precision stage. A

– fibre polarization controller with a prism polarizer selects the transverse-electric polarization state for the cavity mode. The output transmission of the photonic crystal cavity is collected into fiber through a coupling lens, an optical isolator, and then into a New Focus (Model 1811) detector, prior to an electronic spectrum analyzer (Agilent N9000A) and time-domain digital oscilloscope (Tektronix TDS 7404) characterization and statistical analysis..

**Numerical simulations:** The coupled equations (1)-(4) are numerically solved with the fourth-order Runge-Kutta algorithm. The time discretization is set as 10 ps and each simulated temporal waveform contains $10^7$ data points (100 μs). The simulated RF spectrum is calculated with the fast Fourier transform (FFT) method, which is a discrete Fourier transform algorithm to rapidly convert a signal from its time domain to a representation in the frequency domain. In frequency domain, we can easily get the spectral characteristics of the signal. The long time span of the temporal waveform (at 100 μs) is also necessary for resolving the 25 kHz spectral features and converging in the subsequent statistical analyses.

**Data availability.** The authors declare that the data supporting the findings of this study are available within the paper and its Supplementary Information files.

heterostructured photonic crystal cavities. *Appl. Phys. Lett.* **104**, 061104 (2014).
31. Johnson, T. J., Borselli, M. & Painter, O. Self-induced optical modulation of the transmission through a high-Q silicon microdisk resonator. *Opt. Express* **14**, 817-831 (2006).
32. Sprott J. C. *Chaos and Time Series Analysis* (Oxford University Press, 2003).
33. Ott E. *Chaos in Dynamical Systems* (Cambridge University Press, 2002).
34. Grassberger, P., Procaccia, I. Measuring the strangeness of strange attractors. *Phys. D: Nonlinear Phenom.* 9, 189-208 (1983).
35. Schuster, H. G. *Deterministic Chaos: An Introduction* 3rd edn (Wiley, 1995).
36. Grassberger, P., Procaccia, I. Characterization of strange attractors. *Phys. Rev. Lett.* **50**, 346-349 (1983).
37. Strogatz, S. H. *Nonlinear dynamics and chaos* (Addison-Wesley, 1994).
38. Grassberger, P., Procaccia, I. Estimation of the Kolmogorov entropy from a chaotic signal. *Phys. Rev. A* **28**, 2591-2593 (1983).
39. Gilbert T. & Gammon, R. Stable oscillations and Devil's staircase in the Van der Pol oscillator. *Int. J. Bifurcation Chaos* **10**, 155-164 (2000).
40. Lin F. Y. & Liu, J. M. Harmonic frequency locking in a semiconductor laser with delayed negative optoelectronic feedback. *Appl. Phys. Lett.* **81**, 3128-3130 (2002).
41. Jiang W. C., Lu, X., Zhang J. & Lin, Q. High-frequency silicon optomechanical oscillator with an ultralow threshold. *Opt. Express* **20**, 15991-15996 (2012).
42. Zhang M. *et al.* Synchronization of micromechanical oscillators using light. *Phys. Rev. Lett.* **109**, 233906 (2012).

**Acknowledgements**

The authors acknowledge discussions with Eli Kinigstein, Jing Dong, Jaime Gonzalo Flor Flores, and Xingsheng Luan, and with Jiangjun Zheng on the initial design layout and measurements. This material is supported by the Office of Naval Research (N00014-14-1-0041), the National Science Foundation of China (11474233, 61372139, 61571372，61672436), the Chongqing College Teacher Fund (102060-20600512), and the Air Force Office of Scientific Research under award number FA9550-15-1-0081.

**Author contributions**

J.G.W., Y.J.H., and H.Z. performed the measurements, J.G.W., S.W.H., J.H.Y., and C.W.W. performed the numerical simulations and designed the layout, M.B.Y., G.Q.L, and D.L.K. performed the device nanofabrication, and J.G.W., S.W.H., J.M.L., S.K.D. and C.W.W. discussed and put together the manuscript with contributions from all authors.

**Additional Information**

The authors declare no competing financial interests. Supplementary Information accompanies this paper online. Correspondence and requests for materials should be addressed to S.K.D. and C.W.W.

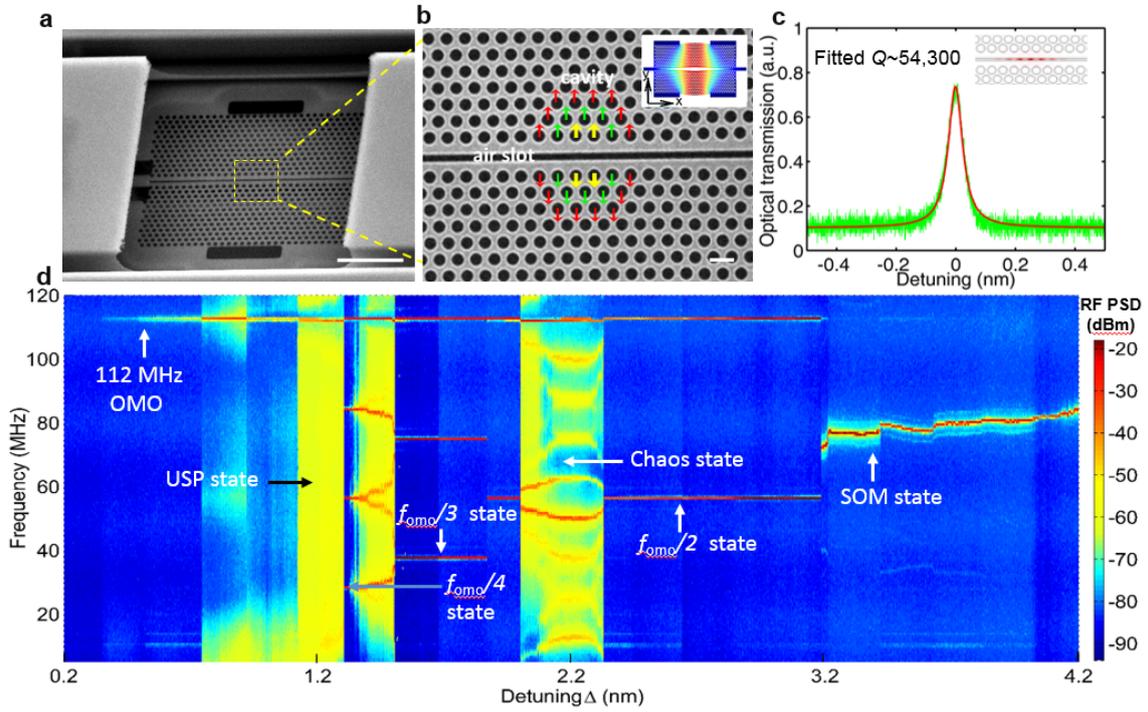

**Figure 1 | Observations of dynamical chaos in mesoscopic optomechanical cavities.** (**a**) Scanning electron micrograph of the optomechanical cavity. Scale bar: 5 μm. (**b**) Zoom-in of 120 nm slot cavity with localized resonant mode formed by perturbed neighbouring holes at the cavity centre, with amplitude displacements denoted by the coloured arrows (yellow: 15 nm; green: 10 nm; red: 5 nm). The lattice constant is 500 nm and the ratio between hole radius and lattice constant is 0.34. Scale bar: 500 nm. Inset: finite-element model of the fundamental mechanical mode field. (**c**) Measured optical transmission spectrum with a cold cavity loaded quality factor $Q$ of 54,300 under low injection power and centred at 1572.8 nm. Inset: $|E|^2$ field distribution of the fundamental optical resonance. (**d**) 2D radio frequency (RF) spectral map illustrating the evolution of nonlinear and chaotic dynamics, detailed as OMO (optomechanical oscillation) state - USP (unstable pulse) state - $f_{omo}/4$ state - $f_{omo}/3$ state - chaos state - $f_{omo}/2$ state - SOM (self-induced optical modulation) state, under controlled laser-cavity detuning $\Delta$ and at 1.26 mW injection power.

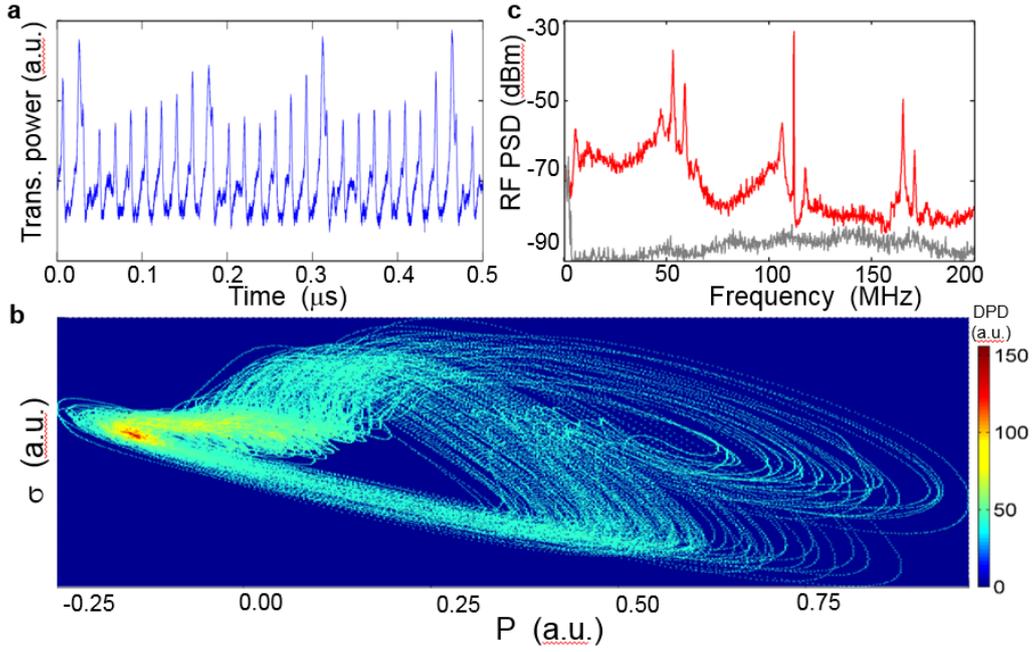

**Figure 2 | Frequency-time characterization of the chaos.** (**a**) Raw temporal waveform of chaotic output. (**b**) Corresponding phase portraits of noise reduced temporal waveform, where the colour evolution from cyan blue – orange - red is proportional to the data point density (DPD) in the measured temporal orbit. (**c**) Corresponding measured radio frequency (RF) power spectral density (PSD). The grey curve is the reference background noise floor.

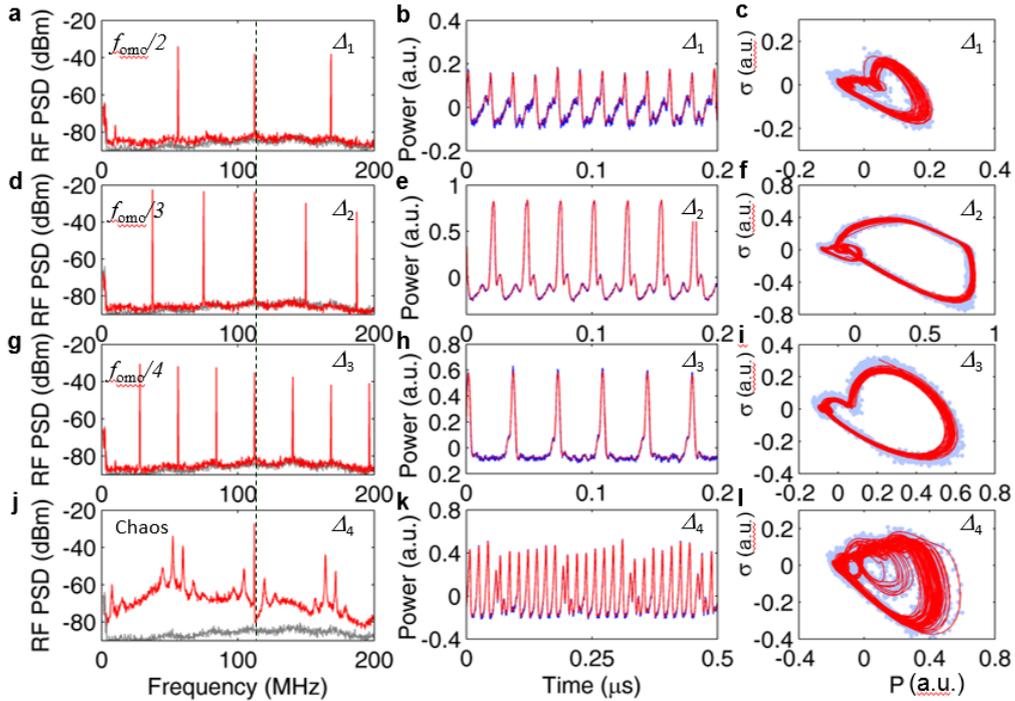

**Figure 3 | Dynamical states under controlled drive conditions.** (**a-c**) the $f_{omo}/2$ state ($\Delta_1 \approx$

2.406 nm), (**d-f**) the $f_{omo}/3$ state ($\Delta_2 \approx$ 1.831 nm), (**g-i**) the $f_{omo}/4$ state ($\Delta_3 \approx$ 1.394 nm) and (**j-l**) the chaos state ($\Delta_4 \approx$ 2.285 nm) respectively. The curves (**a**, **d**, **g** and **j**) are the measured RF power spectral density (PSD) where the grey curves are the background noise floor. Notice the subharmonics have the background at the noise floor. The curves (**b**, **c**, **e**, **f**, **h**, **i**, **k** and **l**) are the temporal waveforms and orbital phase portraits, where the blue dots are the measured raw data and the solid red curves are the noise-reduced orbital trajectories.

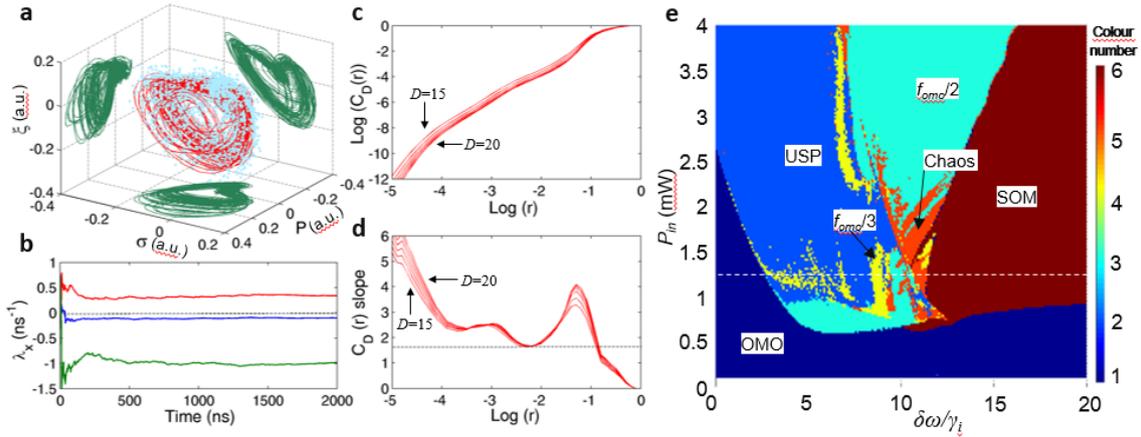

**Figure 4 | Chaos identification and regime distribution map.** (**a**) Measured three-dimensional portrait in phase space. Blue dots are the measured raw data, while the solid red curve is the reconstructed trajectory. The three green phase portraits are projections of the 3D portrait onto the phase planes. (**b**) Calculated Lyapunov exponents (LEs) spectrum. The curves converge to the LE values as $\lambda_1 \approx$ 0.329 ns$^{-1}$, $\lambda_2 \approx$ -0.087 ns$^{-1}$ and $\lambda_3 \approx$ -0.946 ns$^{-1}$. (**c**) Logarithmic plots of correlation integral $C_D(r)$ versus sphere radius $r$ based on the Grassberger-Procaccia algorithm. (**d**) Slope of the correlation integral versus sphere radius $r$. A clear plateau on the slope of the correlation integral is observed, and marked with the horizontal dashed line. The correlation dimension $D_2$ is estimated at $\approx$ 1.67. In panel **c** and **d**, the lines denote the vector size $D$ from 15 to 20 in integer steps. (**e**) Dynamical distribution map based on the numerical modeling. Different colors denote with different dynamical states, including OMO (optomechanical oscillation) state, USP (unstable pulse) state, $f_{omo}/3$ state, $f_{omo}/2$ state, chaos state and SOM (self-induced optical modulation) state. The OMO state, USP state, $f_{omo}/3$ state and $f_{omo}/2$ state denote the periodic and low entropy dynamical regimes; the chaos state and SOM state denote the high entropy dynamical regimes. The horizontal axis is the

normalized laser-cavity detuning $\delta\omega/\gamma_i$ and the vertical axis is the injected optical power $P_{in}$. The dashed white horizontal line is an example corresponding to the injected power level in the measurement.